\documentstyle[aps,preprint]{revtex}

\begin{document}
\draft
\title{Universal and non-universal features of glassy relaxation
in propylene carbonate}

\author{W.~G\"otze and Th.~Voigtmann}
\address{
Physik-Department, Technische Universit\"at M\"unchen, 
D-85747 Garching, Germany} 

\date{received ... November 1999}

\maketitle

\begin{abstract} 
It is demonstrated that the susceptibility spectra of supercooled
propylene carbonate as measured by
depolarized-light-scattering, dielectric-loss, and incoherent quasi-elastic
neutron-scattering spectroscopy within the GHz window
are simultaneously described by the
solutions of a two-component
schematic model of the mode-coupling theory (MCT) for the
evolution of glassy dynamics.
It is shown that the universal $\beta$-relaxation-scaling laws,
dealing with the asymptotic behavior of the MCT solutions,
describe the qualitative features of the calculated spectra.
But the non-universal corrections to the scaling laws render it impossible to
achieve a complete quantitative description using only the
leading-order-asymptotic results.
\\
\end{abstract}

\section{Introduction}

In this paper the evolution of structural relaxation as observed upon
cooling the van-der-Waals liquid propylene carbonate (PC) from above the
melting temperature ($T_m=218\,\mathrm K$) to the glass-transition
temperature ($T_g=160\,\mathrm K$) will be analyzed. It will be shown
that the spectra, as measured within the four-decade frequency window
below $800\,\mathrm GHz$ by depolarized-light-scattering, by dielectric-loss,
and by neutron-scattering spectroscopy can be quantitatively described by
the solutions of a two-component schematic model of the mode-coupling theory
(MCT), where the drift of the various spectral features over several orders
of magnitude due to temperature changes can be fitted by smooth
variations of the model parameters. The results of the data fits will be
used to demonstrate in detail which features can be explained by the
universal $\beta$-relaxation-scaling laws of the
asymptotic MCT-bifurcation dynamics,
and which are caused by either
preasymptotic corrections to this scaling or by crossover phenomena to
microscopic oscillatory motion.

Glassy PC spectra within the full GHz window have first been studied
by Du et~al.\ \cite{Du94} using
depolarized-light-scattering spectroscopy. It was shown that
the data can be interpreted with the universal laws predicted by MCT.
In its basic version, which is
also referred to as the idealized MCT, this theory implies an ideal
liquid-glass transition at a characteristic temperature $T_c$. In an
extended version, $T_c$ marks a crossover from the high-temperature
regime, where the dynamics is dominated by non-linear-interaction effects
between density fluctuations, to a low-temperature regime, where the dynamics
deals with activated-hopping transport in an effectively frozen system.
For temperatures $T$ near $T_c$, the MCT equations can be solved
by asymptotic expansions for the so-called $\beta$-relaxation regime.
This results in formulas for universal features
of the MCT dynamics as reflected in the appearance of dynamical scaling
laws, power-law-decay processes, and in algebraically diverging time scales.
The different anomalous exponents and also the $\beta$-relaxation-master
functions are determined by a system-dependent number which is called the
exponent parameter $\lambda$ \cite{Goetze92}.
The data analysis of Ref.~\cite{Du94}
suggested $T_c\approx187\,\mathrm K$ and $\lambda\approx0.78$. Relaxation
curves measured for PC within the pico-second window in solvation-dynamics
studies \cite{Ma96} and dielectric-loss spectra determined within the
GHz window \cite{Lunkenheimer97d,Lunkenheimer97,Schneider99} have
also been analyzed
with the MCT-scaling-law formulas using parameters $T_c$ and $\lambda$
consistent within the experimental uncertainties with the values cited above.
The critical temperature for PC has first been determined to $T_c\approx180
\,\mathrm K$ \cite{Boerjesson90} by interpreting the $\alpha$-relaxation time
for density fluctuations measured by neutron-scattering spectroscopy
with the MCT-power-law prediction for this quantity. A similar analysis
of the viscosity \cite{Du94,Boerjesson90} suggests a value of $T_c$ near
$190\,\mathrm K$.
The effective Debye-Waller factor for the elastic modulus has been measured
for PC by Brillouin-scattering spectroscopy \cite{Elmroth92}.
Interpreting this quantity with the asymptotic formula of the idealized MCT,
a critical temperature considerably
higher than $190\,\mathrm K$ has been suggested. However, since the
data interpretation is not compelling \cite{Du94},
this finding cannot be
considered to be a falsification of the $T_c\approx187\,\mathrm K$ result.
Thus one could conclude, that MCT describes some essential features of the
glassy dynamics of PC qualitatively correct, a statement which also holds
for a series of other glass-forming systems \cite{Goetze99}.

In order to arrive at a more stringent assessment of MCT,
Wuttke et~al.\ \cite{Wuttke99} have re-examined the above cited PC data for
$T>T_c$. In addition, they have studied incoherent-neutron-scattering
spectra $S(q,\omega)$ for a two-decade window in frequency $\omega$ and for
wave vectors $q$ between $0.7$ and $2.3\,{\mathrm{\AA}}^{-1}$.
The data exhibited
the predicted factorization in a $q$-dependent but $\omega$-independent
amplitude $h_q$, and a $q$-independent term describing the frequency
and temperature variation:
$S(q,\omega)\propto h_q\chi''(\omega)/\omega$. The susceptibility spectrum
$\chi''(\omega)$ showed the subtle dependence
on $\omega$ and on $(T-T_c)$ predicted by the MCT-scaling laws for the
$\beta$-process, provided
$T_c\approx182\,\mathrm K$ and $\lambda\approx0.72$
was chosen. These parameters are marginally compatible
with the values found in the above cited earlier work on PC. The
depolarized-light-scattering spectra have been remeasured within the
$\beta$-relaxation window for $T>T_c$.
The spectrometer used in Ref.~\cite{Wuttke99} incorporated several improvements
over the one used in the original study \cite{Du94}, resulting in improved
signal-to-noise ratios. Furthermore, the use of a narrow-band interference
filter eliminated the possibility of higher-order transmission effects which
have recently been recognized as a potential source of artifacts
\cite{Surovtsev98,Barshilia99}.
But the new spectra agree with the old ones within the error bars of the
latter. The remeasured spectra could be fitted convincingly
with the universal asymptotic
results using the newly found values for $T_c$ and $\lambda$. It was shown
in addition that also the solvation-dynamics results \cite{Ma96} and
the dielectric-loss spectra \cite{Lunkenheimer97d,Lunkenheimer97,Schneider99}
could be fitted within the same frame using the new values
for $T_c$ and $\lambda$.
Actually, the new fit to the dielectric-loss data \cite{Wuttke99}
is more convincing than the original one
\cite{Lunkenheimer97d,Lunkenheimer97,Schneider99},
since the fit interval expands with decreasing $(T-T_c)$,
as requested by MCT. The size of the $(T-T_c)$ interval and the window for
the frequency  where leading-order-asymptotic results describe the
MCT-bifurcation dynamics, depend on the probing variable
\cite{Franosch97,Fuchs98}. It was assumed in
\cite{Wuttke99} that the range of validity of the asymptotic
analysis is smaller for the dielectric-loss spectra than for the
light-scattering spectra. It also had to be anticipated that preasymptotic
corrections can account for a $35\%$ offset of the $\beta$-relaxation-time
scale of the neutron-scattering data relative to
the one for the light-scattering data.

To corroborate the cited MCT interpretations of glassy PC spectra, the
previous work shall in this paper be extended in three directions.
First, the $\alpha$-relaxation peaks will be included in the analysis,
so that the low-frequency limit for the fit interval can be decreased to
$1\,\mathrm GHz$ or lower. Thereby the crossover from $\alpha$- to
$\beta$-relaxation and the non-universal $\alpha$-peak shapes
can be described as well. Second, the crossover from relaxation
to vibrational dynamics will be included in the analysis, so that the
high-frequency limit for the fit interval can be increased by about a factor
of four.
Third, an extended form of the
MCT instead of the idealized one
will be used, so that the spectra for depolarized-light scattering and
dielectric loss for $\omega\ge1\,\mathrm GHz$ can be described also for
temperatures below $T_c$. The specified goals will be achieved
by studying the full solutions of an MCT model.

The paper is organized as follows:
In Sec.~\ref{model-def}, the basic formulas for the
schematic model to be used will be summarized, and then
(Sec.~\ref{exp-analysis}) the experimental data sets are fitted
using this model with smoothly drifting parameters.
After a short introduction to the necessary equations
for the asymptotic analysis for the model (Sec.~\ref{beta-intro}),
the $\beta$-scaling laws are tested against the data in
Sec.~\ref{beta-analysis}.
In Sec.~\ref{de-modulus}, it will be shown that for the
studied model a properly defined
dielectric modulus is more suited for a description by scaling laws
than the dielectric function.
Section~\ref{conclusions} presents some conclusions.

\section{A schematic mode-coupling-theory model}
\label{model-def}

The idealized MCT is based on closed equations
of motion for the auto-correlation functions of the density fluctuations
$\phi_q(t)$, which are positive definite functions of time $t$, depending
on the wave vector modulus $q$ \cite{Bengtzelius84}. The extended MCT also
includes couplings of the density correlators $\phi_q(t)$
to the auto-correlation
functions for the currents \cite{Goetze87}. The general equation of motion
expresses the density correlator in terms of relaxation kernels. It is
formulated most transparently with Laplace-transformed quantities.
For the latter,
the convention $F(z)=i\int_0^\infty \exp(izt)F(t)\,dt$ with complex
frequency $z$, and $F(\omega)
=F'(\omega)+i F''(\omega)$ for $z=\omega+i0$ will be used.
\begin{mathletters}\label{fullmct}
\begin{eqnarray}\label{fullmct-a}
  &\displaystyle\phi_q(z)=\frac{-1}{z+C_q(z)}\;,&\\
  &\displaystyle C_q(z)=N_q(z)
              -\frac{\Omega_q^2}{z+M_q^{\text{reg}}(z)
                     +\Omega_q^2m_q(z)}\;.&
\end{eqnarray}
Here, $\Omega_q$ denotes a characteristic frequency given by the thermal
velocity $v$ and the static structure factor $S_q$: $\Omega_q^2=q^2v^2/S_q$.
The general current-flow kernel $C_q(z)$ describes density-fluctuation
decay via two parallel channels. Phonon-assisted hopping is given by
$N_q(z)$. The relaxation due to nonlinear interactions of density
fluctuations is described by a force-fluctuation kernel which consists of
a sum of a regular term $M_q^{\text{reg}}(z)$ and a mode-coupling term
$m_q(z)$. The former deals with normal-liquid dynamics, and the latter
with the slow motion caused by the cage effect. It is obtained as a
polynomial ${\mathcal F}_q$ of the density correlators $\phi_q(t)$:
\begin{equation}\label{fullmct-m}
  m_q(t)={\mathcal F}_q\left[\phi_q(t)\right]\;. \end{equation}
\end{mathletters}
The coefficients of the polynomial are non-negative; they are given by the
equilibrium structure and hence depend smoothly on external control
parameters like temperature $T$. Systematic studies of the kernels $N_q(z)$
and $M_q^{\text{reg}}(z)$ are not available. The theory shall
be simplified by Markov approximations of these quantities:
$M_q^{\text{reg}}(z)=i\nu_q$, $N_q(z)=i\Delta_q$. The friction
constants $\nu_q\ge0$ and hopping coefficients $\Delta_q\ge0$ shall be
treated as model parameters, which depend smoothly on $T$.

Equations~(\ref{fullmct}) can exhibit bifurcation singularities.
Generically, if as a single control parameter the temperature
is considered, the singularity occurs for a critical temperature $T_c$
if all hopping coefficients $\Delta_q$ vanish. If some $\Delta_q\neq0$,
the singularity is avoided. However, for small $\Delta_q$ and small
$|T-T_c|$ the singularity causes an anomalous dynamics: the
glassy dynamics studied by MCT. At the singularity the correlators do not
decay to zero but to a positive value $f_q^c$, which is called the plateau.
It is approached by an algebraic decay law, called the critical decay,
which is specified by an anomalous exponent $a$, $0<a\le1/2$:
\begin{eqnarray}\label{critical}
  \phi_q(t)-f_q^c=h_q(t/t_0)^{-a}+{\mathcal O}\left(t^{-2a}\right);
\nonumber\\
  T=T_c,\quad\Delta_q=0\;. \end{eqnarray}
The quantity $h_q>0$ is called the critical amplitude, and it
can be determined from the mode-coupling
functional ${\mathcal F}_q$ for $T=T_c$. The time scale $t_0$ is determined
by the transient dynamics for $T=T_c$. For $\Delta_q=0$
and small but negative $(T_c-T)$, the correlator falls below the plateau
$f_q^c$ according to the von Schweidler law $\phi_q(t)-f_q^c\propto-t^b
+{\mathcal O}\left(t^{2b}\right)$, characterized by a second anomalous
exponent $b$, $0<b\le1$.
From ${\mathcal F}_q$ for $T=T_c$, one can calculate
the above mentioned exponent parameter $\lambda$, $0<\lambda\le1/2$, which
determines the critical exponent $a$ and the von Schweidler exponent $b$ via
${\Gamma(1-a)^2}/{\Gamma(1-2a)}=\lambda
  ={\Gamma(1+b)^2}/{\Gamma(1+2b)}$.
In the so-called $\beta$-relaxation window, implicitly defined by
$|\phi_q(t)-f_q^c|\ll1$, MCT predicts that the dynamics is in leading order
controlled by merely two smooth functions of $T$:
the separation parameter $\sigma$ and the hopping parameter
$\delta$. The former is determined by ${\mathcal F}_q$, and its zero defines
the crossover temperature $T_c$: $\sigma=C(T_c-T)/T_c+
{\mathcal{O}}\left((T-T_c)^2\right)$.
The latter obeys $\delta\ge0$; generically,
$\delta$ vanishes only if $\Delta_q=0$ for all $q$.
The shape of the correlation functions in the asymptotic regime of the
$\beta$-relaxation window is
fully determined by the exponent parameter $\lambda$;
as can be inferred from Ref.~\cite{Goetze92}
and the original papers cited therein.

Testing the relevance of MCT by comparing the leading-order
results for the $\beta$-relaxation with data
is however hampered by a great difficulty.
Without detailled microscopic calculations one cannot
determine the size of the corrections to the asymptotic formulas,
and therefore their range of validity is not known.
In addition, the optimal choice of $\lambda$,
fixing the shape of the $\log\chi''$-versus-$\log\omega$ graph is tedious
to decide upon and might well depend on
the choice of the fit interval. The difficulty of fixing $\lambda$ from
a $\beta$-relaxation study alone was demonstrated recently for the hard-sphere
system \cite{Sciortino99}. A set of density
correlators $\phi_q(t)$ calculated for various wave vectors and
packing fractions was considered.
A fit to them with the asymptotic predictions for a
significantly wrong $\lambda$ was by a standard fitting procedure not
distinguishable from the correct fits within typical experimental windows.

A different route for data interpretation is based on comparison of the
measured spectra with the complete solutions obtained from schematic MCT
models. This procedure was studied first by Alba-Simionesco et~al.\
\cite{Alba95,Alba96,Krakoviack97}. Schematic models are truncations of the
complete set of Eqs.~(\ref{fullmct}) to a set dealing with a small number
of correlators only. Thus the mathematical complexity of the problem is
reduced considerably.
Alas, the connection of the
mode-coupling-functional coefficients with the microscopic structure gets
lost; the coefficients are
to be treated as fit parameters. The main advantage of this approach is that
one does not rely on the applicability of asymptotic formulas;
one is sure that all results on crossover phenomena and
preasymptotic corrections are logically consistent with the MCT.

The simplest schematic model deals with a single correlator only, which
shall be denoted by $\phi(t)$. The first MCT equation is equivalent to
Eqs.~(\ref{fullmct-a},b) with $q$ indices dropped:
\begin{mathletters}\label{schematic}
\begin{equation}\label{schematic-a}
\phi(z)=\frac{-1}{z+i\Delta-\Omega^2/\left[
  z+i\nu+\Omega^2m(z)\right]}\;. \end{equation}
For the mode-coupling functional, a quadratic polynomial
that can reproduce all
valid values for the exponent parameter $\lambda$ is used \cite{Goetze84}:
\begin{equation}\label{f12-m}
  m(t)=v_1\phi(t)+v_2\left(\phi(t)\right)^2\;.\end{equation}
\end{mathletters}
For $\Delta=0$, ideal liquid-glass transitions occur on a line in the
$v_1$-$v_2$ plane of coupling constants. One can use $\lambda$ to parameterize
this line of critical coupling constants:
\begin{equation}\label{f12-lambda}
v_1^c=(2\lambda-1)/\lambda^2,\quad v_2^c=1/\lambda^2,\quad1/2\le\lambda<1\;.
\end{equation}
Thus this model is specified by two control parameters $(v_1,v_2)$, by
two frequencies $(\Omega,\nu)$ quantifying the transient dynamics, and one
rate $\Delta$ for the activated transport processes. The model has many
non-generic features, and therefore one
cannot expect it to describe a measured spectrum. In the
present paper, the correlator $\phi(t)$
is introduced to mimic in an overall fashion the combined effect of all
structure fluctuations in producing the bifurcation point and the exponent
parameter $\lambda$ of the system.

The dynamics of some probing variable $A$ coupling to density fluctuations
shall be described by a second
correlator, to be denoted $\phi^s_A(t)$. It obeys an equation analogous
to Eq.~(\ref{schematic-a}):
\begin{mathletters}\label{sjoegren}
\begin{equation}\label{sjoegren-a}
  \phi^s_A(z)=\frac{-1}{z+i\Delta^s_A-{\Omega^s_A}^2/\left[
    z+i\nu_A^s+{\Omega^s_A}^2m^s_A(z)\right]}\;. \end{equation}
Again the microscopic dynamics is quantified by two frequencies
referred to as microscopic parameters
$(\Omega_A^s,\nu_A^s)$. The activated relaxation processes
are described by $\Delta_A^s$.
The mode-coupling functional shall be specified by a coupling
to $\phi(t)$ quantified by a single coupling constant $v_A^s$:
\begin{equation}\label{sjoegren-m}
  m^s_A(t)=v^s_A\phi(t)\phi^s_A(t)\;. \end{equation}
\end{mathletters}
It is a peculiarity of this model, that the dynamics of the probing variable
$A$ is influenced by $\phi(t)$ but not vice versa. Thus the position of the
transition is not modified by the introduction of the second correlator nor
is the value of $\lambda$. The model was motivated by Sj\"ogren
\cite{Sjoegren86} for the description of tagged-particle motion in a
glassy environment, and it will be used here in the same context
for the interpretation of the neutron-scattering data.
The MCT for the reorientational dynamics of a non-spherical probe molecule
suggests the same schematic model for the dipole and quadrupole relaxation
\cite{Franosch97c}; an observation that motivates the application of the model
for the description of the dielectric-loss and depolarized-light-scattering
spectra, respectively.
For the incoherent-neutron-scattering cross-section the fit
will be done using the
model parameters for $\phi^s_A(t)$ different for different wave vectors.
For the index $A$ the abbreviations ls, de, and ns for
light scattering, dielectric loss, and neutron scattering, respectively,
will be used.
The specified two-component schematic model has been used earlier
for data interpretation with the restriction to $\Delta=\Delta^s_A=0$.
Depolarized-light-scattering
spectra within the full GHz band have been described for glycerol for all
temperatures above $T_g$ \cite{Franosch97a}, and for ortho-terphenyl for
$T>T_c$ \cite{Singh98}. Ruffl\'e et~al.\ \cite{Ruffle99} were the first to
simultaneously describe glassy spectra for several probing variables $A$.
Within the $\beta$-relaxation regime,
they fitted coherent-neutron-scattering spectra for several wave vectors and
also the longitudinal elastic modulus
for $\mathrm{Na_{0.5}Li_{0.5}PO_3}$.

The single coupling constant $v^s_A$ determines all features of the
structural-relaxation part of the second correlator. Thus, the $\alpha$-peak
strengths, widths, and positions are correlated. These correlations follow
the same pattern as found and explained for the $\alpha$ peaks of the
hard-sphere system \cite{Franosch97,Fuchs98}. Nevertheless, it is not
obvious from the beginning, and thus truly remarkable,
that such a simple model will be sufficient not only to
explain the trends found in the data, but even to reproduce
structural relaxation for PC quantitatively.

Equation~(\ref{schematic-a}) is equivalent to
\begin{eqnarray}\label{schematic-t}
  \ddot\phi(t)+(\Delta+\nu)\dot\phi(t)+(\Omega^2+\Delta\cdot\nu)\phi(t)
\nonumber\\
  +\Omega^2\int_0^t m(t-t')\left[\dot\phi(t')+\Delta\cdot\phi(t')\right]
  \,dt'=0\;,
\end{eqnarray}
to be solved with the initial condition $\phi(t=0)=1$, $\dot\phi(t=0)=-\Delta$.
This equation, together with Eq.~(\ref{f12-m}), is solved numerically with
a similar algorithm as used in the preceding work for
the case $\Delta=0$. Equations~(\ref{sjoegren}) are treated in the same manner,
but $\phi(t)$ has to be used as input for Eq.~(\ref{sjoegren-m}). From the
result for $\phi^s_A(t)$, a Laplace-transformation yields $\phi^s_A(z)$.
The fluctuation-dissipation theorem then determines the dynamical
susceptibility $\chi_A(z)$ of variable $A$:
\begin{equation}\label{susceptibil}
  \chi_A(z)/\chi_A=z\phi^s_A(z)+1\;. \end{equation}
Here, $\chi_A\propto\left<A^2\right>$ is the thermodynamic susceptibility.
In particular, the imaginary part
of Eq.~(\ref{susceptibil})
determines the normalized susceptibility spectrum, $\chi''_A(\omega)/\chi_A
=\omega{\phi^s_A}''(\omega)$, the quantity of main interest in the
following. In our data analysis, $\chi_A$ enters as an additional fit
parameter, which we treat, for the sake of simplicity,
as a temperature-independent normalization constant.

\section{Data analysis}
\label{exp-analysis}

\subsection{Fits to the data}

The result of our fits to the measured
PC spectra are shown by the full lines in Figs.~\ref{fit-ls-de}
and \ref{fit-ns}. Since one cannot expect the schematic model to provide
a description of the microscopic band, the fits have been restricted to
frequencies below $500\,\mathrm GHz$ for the light-scattering and
neutron-scattering spectra. The fit range for the dielectric spectra could
be extended up to $1\,\mathrm THz$. For the neutron-scattering data, a set
of spectra for 3 representative $q$-vectors out of 10 analyzed is shown.
The analyzed $q$-range is
$0.5\,{\mathrm{\AA}}^{-1}\le q\le1.4\,{\mathrm{\AA}}^{-1}$; outside this
range, experimentally accessible frequency windows become too small to gain
meaningful information for MCT parameters.
In Ref.~\cite{Du94}, light-scattering spectra
above $T=250\,\mathrm K$ have been published, but show apparent violation
of $\alpha$ scaling. We were able to fit these curves with the same quality
as the ones shown by assuming a slightly varying static susceptibility
$\chi_{\text{ls}}$, which has the effect of shifting curves up and down in
the log-log plot. These curves were omitted in Fig.~\ref{fit-ls-de} to
avoid overcrowding.

All model parameters should be used as temperature-dependent
fit parameters in our analysis. Within the studied temperature interval,
there are no structural anomalies
reported for PC. Thus, the fits are done with the constraint that the
parameters drift smoothly and monotonously.
In the following part of this section,
the parameters used for the theoretical curves in
Figs.~\ref{fit-ls-de} and \ref{fit-ns} shall be discussed.

One experiences a considerable flexibility in choosing the path
$\left(v_1(T),v_2(T)\right)$ followed by the coupling constants in the
$v_1$-$v_2$-parameter plane for the interpretation of the data
as emphasized earlier \cite{Franosch97a}.
To arrive at an overview of the possibilities for
fitting the many spectra, we started with a first step, where the path was
varied but
biased to some smooth curve. Applying the general theory \cite{Goetze91b}
to Eq.~(\ref{f12-m}), one derives
the formula for the above-mentioned separation parameter $\sigma$,
\begin{equation}\label{sigma-f12}
  \sigma=(1-f^c)\left[(v_1-v_1^c)f^c+(v_2-v_2^c){f^c}^2\right]\;.\end{equation}
In our first step of the analysis, we also force the $v_1$, $v_2$ to
obey the asymptotic linear $(T_c-T)$ dependence of $\sigma$ cited above.
In the second step, this latter restriction is eliminated
and a free fit is started by examining small corrections to the result of
the first step. The thus obtained results also account for an
inevitable uncertainty in the determination of the
experimental temperatures. The fit yields $T_c\approx180\,\mathrm K$,
and $\lambda\approx0.75$, corresponding to $a\approx0.30$ and $b\approx0.56$.
The value for $\lambda$ is between the values reported in Refs.~\cite{Du94}
and \cite{Wuttke99} and falls within the error bars of both.
The linear interpolation of the found $\sigma$ versus $T$ values gives
$\sigma=C(T_c-T)/T_c$ with $C\approx0.069$. The
found distribution of $(v_1,v_2)$ points is shown in the upper part of
Fig.~\ref{path}. Upon lowering $T$, both $v_1$ and $v_2$ increase, which
is consistent with the physical reasoning of the system's
mode-coupling coefficients becoming larger at
lower temperatures. The lower diagram in Fig.~\ref{path} demonstrates that
the asymptotic formula for $\sigma$ is well obeyed
for $150\,{\mathrm K}\le T\le285\,{\mathrm K}$. It should be stressed that
the glass-transition line is just crossed by a regular drift, i.~e.\ there
is no accumulation of $(v_1,v_2)$ points close to it. This demonstrates how
the critical phenomena predicted by the MCT originate from the mathematical
structure of its equations of motion. In particular, the schematic model
illustrates that within MCT no subtle $q$-interferences
or hydrodynamic phenomena are responsible for the glass-transition
dynamics.

The fitted mode-coupling coefficients $v^s_A(T)$ for the light-scattering
and dielectric data, and the corresponding coefficients $v_{\text{ns}}^s(q,T)$
for the neutron-scattering experiment are shown in Fig.~\ref{param-vs}. Again,
we find monotonically increasing couplings with decreasing temperature.
The coupling coefficients $v_{\text{ns}}^s(q)$ describing the
incoherent-neutron-scattering data are decreasing with increasing $q$.
This is equivalent to the plateau values $f^{s,c}_q$ decreasing with
increasing $q$, which agrees qualitatively
with the findings for incoherent-neutron-scattering results
discussed within the microscopic MCT \cite{Fuchs98}.

The parameters $\Omega^s_A$, $\nu^s_A$ which specify the transient dynamics
of $\phi^s_A$ are shown in Fig.~\ref{param-micr}. The results from the
neutron-scattering analysis reflect the behavior $\Omega^s_{\text{ns}}(q,T)
\propto q\cdot\sqrt{T}$ to a good approximation, which is in agreement with the
result of the microscopic theory. But drawing more conclusions from the
microscopic parameters would be over-interpreting the model. They
are shown here mainly to demonstrate that there are no abnormal
variations occurring. We find much larger uncertainties for the microscopic
fit parameters $\Omega$, $\nu$, $\Omega^s_A$, $\nu^s_A$, than for those
parameters $v_1$, $v_2$, and $v^s_A$,
ruling the structural-relaxation part of the spectra. In particular,
it was possible to
use for the parameters which specify the transient of the first correlator
$\phi(t)$ temperature independent values $\Omega=1\,\mathrm THz$ and
$\nu=0\,\mathrm THz$.

The hopping coefficient $\Delta$ in Eq.~(\ref{schematic-a}) determines the
position of the susceptibility minimum below $T_c$. This minimum cannot be
seen in the light-scattering data, thus the chosen values are not
unambiguously determined. The light-scattering spectra in the upper panel
of Fig.~\ref{fit-ls-de} are fitted with the hopping parameter $\Delta^s_A$
for the second correlator ignored: $\Delta^s_{\text{ls}}=0$. The fits to the
dielectric-loss spectra in the lower panel of Fig.~\ref{fit-ls-de} are done
with a non-vanishing $\Delta^s_{\text{de}}$. For the whole temperature
range investigated, $\Delta(T)$ can be assumed to follow an Arrhenius law,
$\Delta(T)\propto\exp(-E_A/T)$, which would be expected for thermally activated
hopping over barriers. Fig.~\ref{param-delta} shows the values used for the
fit. Although $\Delta$ increases
by an order of magnitude, the calculated curves for temperatures
higher than $190\,\mathrm K$ show no influence from hopping effects on
the spectra. This is
demonstrated in Fig.~\ref{delta-d0cmp}.
The irrelevance of the increasing hopping coefficients $\Delta_q$ for
temperatures increasing above $T_c$ can be understood on the basis of a
discussion of the asymptotic formulas \cite{Fuchs92}.
It is the reason, why the idealized theory can be used for
data analysis for $T$ sufficiently larger than $T_c$. In the analyzed
neutron-scattering experiment, the dynamical window and the studied
temperature intervals are too
small to investigate hopping effects, and therefore the curves
in Fig.~\ref{fit-ns} are calculated with $\Delta^s_{\text{ns}}=0$.

Above $T_c$, the spectra including hopping show deviations
from the idealized ones only for small $T-T_c$.
Below $T_c$, the crossover to the white-noise
spectrum is suppressed, and a minimum occurs as hopping starts to be the
dominant relaxation effect. Because of the insensitivity of the main body of
the analyzed data to choices of $\Delta$, the activation energy cannot be
determined very precisely from the fit; the upper straight line in
Fig.~\ref{param-delta} corresponds to $E_A=811\,\mathrm K$.
This value is in reasonable agreement with the one found in an earlier
asymptotic analysis \cite{Du94}.
Dielectric-loss
spectra show hopping-induced minima at higher frequencies than the
light-scattering spectra, and this we have accounted for by introducing
a second hopping parameter $\Delta^s_{\text{de}}$ there. In a similar way,
$(\Omega^s_{\text{de}}/\Omega)^2\Delta^s_{\text{de}}(T)$
follows an Arrhenius law and has no influence
on the spectra above $T_c$; this second hopping term has already been
included in the comparison studied in Fig.~\ref{delta-d0cmp}. Here, the
activation energy is of the order of $2000\,\mathrm K$, which makes the
result more striking, since
$(\Omega^s_{\text{de}}/\Omega)^2\Delta_{\text{de}}^s$ is allowed to vary over
three orders of magnitude.
In both cases, activation energies as well as the prefactors
are of reasonable magnitude.
It should be stressed that, although the treatment
of hopping by a frequency-independent $\Delta$ is rather crude, the
resulting frequency range in which the schematic model gives a good fit to
experimental data, is enlarged by about one decade for $T<T_c$ relative
to the fit interval which can be treated by the idealized-MCT model.

In the measurements of the dielectric functions,
information on both the imaginary and the
real part of $\varepsilon(\omega)=\varepsilon'(\omega)
+i\varepsilon''(\omega)$ have been obtained \cite{Schneider99}.
The fit to the $\varepsilon''$
data shown above was performed using $\varepsilon''(\omega)=
4\pi\chi''_{\text{de}}(\omega)=
4\pi\chi_{\text{de}}\omega{\phi^s_{\text{de}}}''(\omega)$, thus obtaining
the proportionality factor $\varepsilon_0=4\pi\chi_{\text{de}}$ as a
by-product. Then, the real part is given by
$\varepsilon'(\omega)-\hat\varepsilon=
\varepsilon_0\cdot(1+\omega{\phi^s}'_{\text{de}}(\omega))$. The new parameter
$\hat\varepsilon$ has to be determined by shifting the curves, and it can
differ from $\varepsilon_\infty=1$ in both directions: The liquid exhibits
microscopic oscillations, which contribute to $\varepsilon'(\omega)$ as some
shift $\Delta\varepsilon_{\text{micr.}}^{\text{exp}}$ with respect to
$\varepsilon_\infty=1$ for the structural
part of the response function. The schematic model uses a single damped
oscillator, giving some $\Delta\varepsilon_{\text{micr.}}^{\text{fit}}$,
which may be either too small or too large. Depending on the temperature,
we find values of
$\hat\varepsilon=\varepsilon_\infty+\left(
 \Delta\varepsilon_{\text{micr.}}^{\text{exp}}
-\Delta\varepsilon_{\text{micr.}}^{\text{fit}}\right)$ between $3$ and $-1$,
which are of reasonable magnitude.
Figure~\ref{fit-de.re} shows
the result of testing our fit against the accordingly shifted real part of the
measured dielectric function.
It is clear from the theory that the real and imaginary parts of the
calculated curves are connected by Kramers-Kronig relations. But for the
experiment, both quantities have to be regarded as almost independent
data sets, since the measurements are restricted to a finite
frequency range. Thus, figure~\ref{fit-de.re} provides more than just a
different view on the fit shown in Fig.~\ref{fit-ls-de}, and it is an
important point that the real-part data can be fitted with the schematic
model as well, introducing only one additional fit parameter $\hat\varepsilon$.
In the minimum region of the
spectra, we find this to be confirmed, and for higher $T$, the
$\alpha$-relaxation step can be described by the schematic model, too. The
discrepancies for the $\alpha$ peak in the glass are the analogue to what
can be seen in the $\varepsilon''$ fit. Similar observations hold for the
high-frequency dynamics, where one has to notice in addition, that experimental
error bars are relatively large for frequencies above $300\,\mathrm GHz$.
A slightly better fit of the $\varepsilon'$ data could have been achieved by
allowing the static susceptibility $\chi_{\text{de}}$ to vary with
temperature. This possibility is not examined here, since the shift is only
small, and since we do not want to introduce assumptions on the $T$-dependence
of the static quantity $\varepsilon_0$.
 
\subsection{Summary of the Data Analysis}

Glass-forming liquids exhibit temperature-sensitive spectra for frequencies
well below the band of microscopic excitations. These precursors of the
glass transition are referred to as structural-relaxation spectra.
The full lines in Fig.~\ref{fit-ls-de} and \ref{fit-ns} demonstrate, that
the evolution of structural relaxation of PC, including the crossover to
the microscopic regime, is described well by a schematic
MCT model. The description holds for all spectra obtained by the
depolarized-light-scattering spectrometer; in this case it deals with the
three-decade dynamical window between $0.3$ and $500\,\mathrm GHz$, and it
accounts for the change of the spectral intensity by a factor of $10^3$
if the temperature is shifted between the glass transition $T_g$ and
$30\,\mathrm K$ above the melting temperature $T_m$.
It accounts for the measured $\alpha$-peak-maximum
shift by a factor of $10$ if $T$ is changed by $30\,\mathrm K$.
A similar statement holds for the description of the dielectric-loss
spectra, where the $\alpha$-peak shift from $40\,\mathrm GHz$ down to
$0.02\,\mathrm GHz$ is described. This shift is caused by a temperature
decrease from $293\,
{\mathrm K}=T_m+75\,\mathrm K$ to $243\,\mathrm K$.

Between the $\alpha$ peak and the vibrational excitation peak near
$1\,\mathrm THz$, the susceptibility spectra
in Figs.~\ref{fit-ls-de} and \ref{fit-ns} exhibit a minimum at some
frequency $\omega_{\text{min}}$. It shifts to smaller frequencies
as the temperature is lowered, but less than the $\alpha$-peak position.
Its intensity $\chi_{\text{min}}
=\chi''(\omega_{\text{min}})$ exceeds the white-noise spectrum one would
expect for the dynamics of normal liquids by more than two orders of
magnitude. Such white noise would yield susceptibility spectra varying
linearly with frequency, $\chi''_{\text{wh.n.}}(\omega)\propto\omega$,
as is indicated by the dashed lines in Fig.~\ref{fit-ls-de}.
These anomalous minima are also treated properly by the model.

Neutron-scattering data are available for a series of wave vectors $q$,
and hence the dynamics is probed on various length scales.
The $q$ dependence is in the schematic model described by that of the
coupling coefficient $v^s_{\text{ns}}(q)$.
The data description in Fig.~\ref{fit-ns} is possible using a $q$
dependence in qualitative agreement with the results expected from the
microscopic theory of simple systems.

It appears nontrivial that the used schematic model can deal with the
mentioned spectra of PC. The success of the fits indicates that the studied
glassy dynamics is rather insensitive to microscopic details of the
systems. Apparently the evolution of glassy dynamics within the GHz window
reflects, above all, only quite general features of the nonlinear-interaction
effects, which can also be modelled by simple truncations of the
full microscopic theory.
These conclusions require some reservation. The explanation of the PC data
by the used model is based on the choice of the model parameters, in
particular on the choice of the drift of all parameters with changes of
temperature, which is documented in Fig.~\ref{path}-\ref{param-delta}. Only
a full microscopic theory can show, whether or not the chosen parameters
are in accord with the fundamental microscopic laws.

Furthermore, it has to be emphasized that the studied model cannot
reproduce the spectra for frequencies below $1\,\mathrm GHz$ if the
temperature is below the critical value $T_c$. Such spectra can be
measured accurately using dielectric-loss spectroscopy, and the lower
panel of Fig.~\ref{fit-ls-de} exhibits some of this data for
$T=173\,\mathrm K$ and $T=183\,\mathrm K$. The lack of success of our
work in handling these spectra is clearly connected with the improper treatment
of hopping processes. It remains unclear at present, whether this
is due to the stochastic approximation, $N_q(z)=i\Delta_q$, or due to
restricting ourselves to a one-component schematic model, or whether
the whole extension of MCT to a theory including hopping transport is
inadequate.

\section{Some asymptotic formulas}
\label{beta-intro}

Let us list some of the asymptotic results for the studied MCT model, which
will be needed below in Sec.~\ref{beta-analysis}.
These results are obtained by straight-forward specialization of the
general formulas discussed in Ref.~\cite{Goetze91b}.
We will focus on the
$\beta$-relaxation regime for $T\ge T_c$, with hopping effects neglected.
A comprehensive discussion
of the asymptotic results
can be found in Ref.~\cite{Franosch97}.

From the full MCT equations (\ref{fullmct}), a leading-order expansion
in $\sqrt{|\sigma|}$ gives rise to the asymptotic predictions for the
intermediate-time window of the $\beta$ relaxation.
A central result is the factorization theorem, $\phi_q(t)-f_q^c=h_qG(t)$,
where the so-called $\beta$ correlator
$G(t)$ is independent of $q$. This result still holds, in the generic case,
for the
tagged-particle density-fluctuation correlator or the
correlator dealing with light scattering or dielectric response:
$\phi^s_A(t)=f_A^{s,c}
+h_A^sG(t)$, with the same $G(t)$ as above.
The Fourier-cosine transform of $G(t)$ is called
the $\beta$ spectrum $G''(\omega)$.
One gets for the normalized susceptibility spectra
\begin{equation}\label{factorisation}
  \chi''_x(\omega)=\omega\phi''_x(\omega)=h_x\chi''(\omega)\;,\end{equation}
where $\chi''(\omega)=\omega G''(\omega)$ is called the
$\beta$-susceptibility spectrum.
Here, the index $x$ denotes either the wave-vector modulus $q$, or
$x=(s,A)$.
The function $G$ depends on $t/t_0$,
$\sigma$, and $\delta$ only: it is uniquely determined by the
exponent parameter $\lambda$ as the solution of the equation
\begin{equation}\label{beta-scaling-eq}
  \sigma-\delta t+\lambda(G(t))^2=
  \frac{d}{dt}\int_0^t G(t-t')G(t')\,dt'\;,
\end{equation}
to be solved with the initial condition $G(t\rightarrow0)=(t/t_0)^{-a}$.
The so-called hopping parameter $\delta$ has to be calculated from $\Delta_q$,
and for the studied model it reads
\begin{equation}\label{delta-f12}
 \delta=\Delta\,{f^c}^2/(1-f^c)\;.
\end{equation}
In this context, the numbers $\Delta_A^s$ only enter as corrections to
scaling.

The plateau values $f_x^c$, and the critical amplitudes $h_x$
can be calculated from the mode-coupling functionals.
In the case of the schematic model studied, the values for the first
correlator are given by $\lambda$:
\begin{equation}\label{f-ampl-f12}
  f^c=1-\lambda,\quad
  h=(1-f^c)\;.
\end{equation}
The relation between the exponent parameter $\lambda$ and
the $\alpha$-peak strength $f^c$ is one of the non-generic features of
that model. For the second correlator, the plateau value and critical
amplitude read
\begin{equation}\label{f-ampl-sjoe}
  f_A^{s,c}=1-\frac{1}{v_A^s f^c},\quad
  h_A^s=\frac{1-f^c}{v_A^s {f^c}^2}\;.
\end{equation}
Changing $v_A^s$, the $\alpha$-peak strength $f_A^{s,c}$
can be varied. Again, these
equations
establish a non-generic relation between the $f_A^{s,c}$ and the $h_A^s$.
In our fits to the neutron-scattering data, a $q$ dependence of
$f_{\text{ns}}^{s,c}$ and $h_{\text{ns}}^s$ can arise only through a
$q$-dependence of the $v_{\text{ns}}^s$. 

From Eq.~(\ref{beta-scaling-eq}) one identifies for the case $\delta=0$
the time scale for the $\beta$-relaxation: $t_\sigma=t_0
|\sigma|^{-1/2a}$. Going over to rescaled times, $\hat t=t/t_\sigma$,
and rescaled frequencies, $\hat\omega=\omega t_\sigma$,
one gets from Eq.~(\ref{factorisation})
the scaling law for the $\beta$-susceptibility spectra
\begin{equation}\label{susc-scaling}
  \chi''_x(\omega)
  =h_x c_\sigma \hat\chi(\hat\omega)\;,
\end{equation}
where $c_\sigma=\sqrt{|\sigma|}$. The master spectrum $\hat\chi$ is
$\sigma$-independent. It is fixed through the exponent parameter $\lambda$,
and thus through the static structure
alone.
For large rescaled frequencies, $\hat\omega\gg1$, one obtains the
critical-power-law spectrum. This extends to all frequencies as
$\sigma\rightarrow0$:
\begin{equation}\label{crit-omega}
  \chi''_x(\omega)=h_x\cdot\sin(\pi a/2)\Gamma(1-a)(\omega t_0)^a,\quad
  T=T_c\;.
\end{equation}

For small rescaled frequencies, one gets the
von Schweidler-law for $\sigma<0$, $\hat\chi(\hat\omega\ll1)
\propto1/\hat\omega^b$,
and thus $\hat\chi$ exhibits a minimum at some frequency
$\hat\omega_{\text{min}}$ with $\hat\chi_{\text{min}}=
\hat\chi(\hat\omega_{\text{min}})$. Due to the scaling law,
Eq.~(\ref{susc-scaling}), the variation of the spectral minima with
temperature is, in the asymptotic region, given by
\begin{equation}\label{min-scaling}
  \omega_{\text{min}}=\hat\omega_{\text{min}}/ t_\sigma,\quad
  \chi_{\text{min}}=\hat\chi_{\text{min}}\cdot c_\sigma,\quad
  \sigma<0\;.
\end{equation}
The point $(\hat\omega_{\text{min}},\hat\chi_{\text{min}})$ is completely
fixed by $\lambda$, and for $\lambda=0.75$ one gets:
$\hat\omega_{\text{min}}=1.733$, $\hat\chi_{\text{min}}=1.221$.

On the glass side, $\sigma>0$, the idealized theory yields for the
$\beta$ correlator for large rescaled times a constant,
$G(\hat t\gg1)=1/\sqrt{1-\lambda}$. Thus the signature of the MCT-fold
bifurcation are
$\sqrt{T_c-T}$ anomalies of the nonergodicity parameters
$f_x=\phi_x(t\rightarrow\infty)$:
\begin{equation}\label{f-sqrt}
  f_x(T)=f_x^c+h_x\sqrt{\sigma/(1-\lambda)},\quad T<T_c\;.
\end{equation}
If the correlators deal with density fluctuations or tagged-particle
densities, the quantity $f_x$ is the Debye-Waller factor or
Lamb-M\"o{\ss}bauer factor, respectively. For $\sigma<0$, corresponding
to $T>T_c$, the long-time limits of the correlators vanish, as is the
case for $T<T_c$ but $\delta\neq0$. But if $\sigma$ and $\delta$ are
sufficiently small, the correlators still exhibit plateaus for times
exceeding the transient scale $t_0$ before the decay towards zero sets in.
The heights of these plateaus are then given by $f_x$ for $T<T_c$,
and by $f_x^c$ for $T>T_c$, then called effective nonergodicity parameters.
The decay from the plateau is the $\alpha$
process, and thus the strength of the $\alpha$ peak in the susceptibility
spectra is given by $f_x$. This also corresponds to
the height of the relaxation step exhibited by the real part of the
susceptibility, when the frequency is shifted through the $\alpha$-peak
window.

The preceding Eqs.~(\ref{factorisation}-\ref{f-sqrt}) establish universality
features of MCT. They provide the basis of a general explanation of the
glassy MCT dynamics by means of features of the spectra
not depending on the specific microscopic properties of a given system.

\section{Scaling law analysis}
\label{beta-analysis}

In this section it shall be studied how well the above calculated MCT
solutions can be described by the MCT-$\beta$-relaxation-scaling laws
summarized in the preceding section.
It has been demonstrated earlier \cite{Franosch97,Fuchs98}, that the range
of validity of these equations can be analyzed by evaluating the
next-to-leading-order corrections. Here, we will study the combined effect
of all corrections due to structural relaxation as well as due to 
vibrational-transient-dynamics effects.
Only the solutions referring to the parameter sets
used in Figs.~\ref{fit-ls-de} and \ref{fit-ns} will be discussed. Thus
the following analysis refers to control parameters and dynamical windows
representative for state-of-the-art experimental studies of the evolution
of glassy dynamics.

\subsection{The critical decay}

Solving the equations of motion for $T=T_c$, $\Delta=0$, and $\Delta^s=0$
for times up to $10^{15}\,\mathrm
ps$, the critical power law, Eq.~(\ref{critical}), was
identified. The common time scale was determined to
$t_0=0.035\,\mathrm ps$. The leading-order result
$\hat\phi_x(t)=(t/t_0)^{-a}$, where $\hat\phi_x(t)=(\phi_x(t)-f_x^c)/h_x$,
is shown in the double-logarithmic representation of
Fig.~\ref{beta-tdom} by straight dash-dotted lines with slope $-a$.
For the two temperatures closest to $T_c$, the full lines in this diagram
exhibit the solutions $\hat\phi_x(t)$.
Dashed lines demonstrate the corresponding
$\beta$ correlators $G(t)$, determined from
Eqs.~(\ref{sigma-f12},\ref{beta-scaling-eq},\ref{delta-f12}).
The approach of the first correlator $\phi(t)$
towards the plateau $f^c$ is well described by the scaling law for
$T=180\,\mathrm K$
and $190\,\mathrm K$. For $T=180\,\mathrm K$ the critical power law
is exhibited within a $1.5$-decade time window for times exceeding $t_c$
with $t_c/t_0\approx300$, while for $t<t_c$ the vibrational transient dynamics
masks the structural relaxation.
In this case, the validity of the critical power law for larger times
is restricted by the onset of hopping effects. Hopping plays no significant
role for the $\beta$ relaxation of $T=190\,\mathrm K$ (compare
Fig.~\ref{delta-d0cmp}). But there, the deviations of $G(t)$ from
the short-time limit
$(t/t_0)^{-a}$ set in already for $t<t_c$. Thus this power law cannot
be identified anymore for distance parameters $\varepsilon=(T_c-T)/T_c$
with $|\varepsilon|\gtrsim0.06$. This scenario
is in semiquantitative agreement with the one discussed
in Ref.~\cite{Franosch98} for the density correlators of a hard-sphere system.
Let us reiterate that the correlator $\phi(t)$ drives the glass transition
for the studied model, but that it is not the quantity measured.

The two lower sets of curves in Fig.~\ref{beta-tdom} show
that the decrease of $\phi^s_{\text{ls}}(t)$
and $\phi^s_{\text{de}}(t)$ towards their plateaus $f^{s,c}_{\text{ls}}$
and $f^{s,c}_{\text{de}}$ respectively is described qualitatively
by the dashed lines, i.~e.\ by the scaling laws. However, there are
remarkable quantitative deviations between the solutions $\hat\phi^s_A(t)$
and their asymptotic form $G(t)$. These appear as if the
amplitude experiences some offset. The reason is that the transient dynamics
influences the correlators $\phi^s_A(t)$ also for times which exceed $t_c$
by up to two orders of magnitude.
This means that the dynamics of the two probing variables is strongly
influenced by oscillations within that window, where the driving correlator
$\phi(t)$ exhibits the $t^{-a}$ law.
Therefore the power law
$\hat\phi^s_A(t)=(t/t_0)^{-a}$ cannot be
identified accurately in the curves shown for $A=\text{ls}$ and $A=\text{de}$.
This is also demonstrated by the straight dash-dotted line in the upper panel
of Fig.~\ref{fit-ls-de}, which represents the asymptotic
low-frequency-susceptibility spectrum at the critical point,
Eq.~(\ref{crit-omega}).

Within the $1.5$-decade window, where the
$\log\hat\phi$-versus-$\log(t/t_0)$ curve for $180\,\mathrm K$ in
Fig.~\ref{beta-tdom} demonstrates the critical-decay asymptote, the graphs
of $\log\hat\phi^s_{\text{ls}}$ and $\log\hat\phi^s_{\text{de}}$ versus
$\log(t/t_0)$ for $180\,\mathrm K$ and $183\,\mathrm K$, respectively,
also appear as nearly straight lines, so that they can be
described very well in this window by some effective power law. One
thus expects an effective power-law spectrum which is described by
Eq.~(\ref{crit-omega}), but with $a$ and $h_A$ replaced by some
$a^{\text{eff}}$ and $h_A^{\text{eff}}$ respectively.
This phenomenon also was observed for
the susceptibility spectra of the hard-sphere system \cite{Franosch98}.
For the light-scattering result, one infers $a^{\text{eff}}<a$
and $h_{\text{ls}}^{\text{eff}}<h^s_{\text{ls}}$.
The dotted line in the upper panel of
Fig.~\ref{fit-ls-de} corroborates this conclusion. It exhibits the solution
for the model evaluated for $T=T_c$ with hopping effects ignored.
This line can be
fitted well between $10^{-5}\,\mathrm THz$
and $10^{-3}\,\mathrm THz$ by an effective
power-law following Eq.~(\ref{crit-omega}) with
$a^{\text{eff}}/a\approx0.92$ and
$h^{\text{eff}}/h\approx0.7$. The crossover from this effective power law to
the asymptotic critical law, Eq.~(\ref{crit-omega}), occurs only at frequencies
around $1\,\mathrm MHz$.

\subsection{The non-ergodicity-parameter anomaly}

Figure~\ref{fig-fs} shows effective nonergodicity parameters of the three
correlators underlying the curves in Fig.~\ref{fit-ls-de}, which
were determined from the plateau heights of
the $\phi_x(t)$-versus-$\log t$ diagrams.
The crosses in the lower panel show
the values deduced in Ref.~\cite{Schneider99} from the step size
of the measured real part of the dielectric function, divided
by the value of $\varepsilon_0$ assumed in the fit to the
susceptibility spectra. Figs.~\ref{fit-ls-de}
and \ref{fit-de.re} demonstrate that the present model describes the dielectric
function of PC reasonably, and so it is not surprising that the calculated
values (dots) reproduce the measured ones (crosses) reasonably well.
The discrepancies between dots and
crosses are anticipated to be mainly due to difficulties in determining the
step size accurately in the experiment, where one carefully has to
eliminate contributions from the $\beta$ relaxation.

Full lines in Fig.~\ref{fig-fs} exhibit the asymptotic laws,
i.~e.\ the values $f_x$ from Eq.~(\ref{f-sqrt}) for $T<T_c$ and the constant
$f_x^c$ for $T\ge T_c$.
Figure~\ref{fig-fs}(a) demonstrates that the $60\%$ variation of the effective
non-ergodicity parameter $f$ of the first correlator is described well by
the asymptotic formula. This holds for temperatures down to $T_g$. On the
other hand, the results for the light scattering and for the dielectric
response do not exhibit the asymptotic behavior; there is no evidence for
the $\sqrt{T_c-T}$ anomaly at all to be noticed in the data.
There are two reasons for this finding.
The obvious one reflects the large size of $f^{s,c}_A$, i.~e.\ it
results from the observation that the $\alpha$ peaks of the susceptibility
dominate over the remaining susceptibility spectrum (compare
Fig.~\ref{fit-ls-de}). Equation~(\ref{f-sqrt}) for probing-variable
correlator is equivalent to
\begin{mathletters}
\begin{equation}\label{f-fs}
(1-f^s_A)=(1-f^{s,c}_A)-h^s_A\sqrt{\sigma/(1-\lambda)}\;.
\end{equation}
Since $(1-f^s_A)$ and $h^s_A$ are positive and $(1-f^{s,c}_A)$ is less than
$0.1$ for the two correlators discussed, the whole $\sqrt{T_c-T}$ effect is
below $10\%$. Therefore it is difficult to separate the $\sqrt{T_c-T}$ anomaly
from the scatter of the data. The less obvious reason results from the
smooth but appreciable temperature drift found for the coupling
coefficient $v^s_A$ (compare Fig.~\ref{param-vs}). This coupling determines
the non-ergodicity parameters of the second correlator of the schematic
model in terms of the parameter $f$: $(1-f_A^s)=1/(v_A^s f)$.
The square-root singularity is due to that in $f$, and expanding
$(1/f)$ one reproduces Eq.~(\ref{f-fs}), but with effective terms
\begin{eqnarray}\label{f-fs-eff}
(1-f^{s,c}_A)^{\text{eff}}=R_A(1-f^{s,c}_A),\quad
h_A^{\text{eff}}=R_A h^s_A,\nonumber\\
R_A(T)=v^{s,c}_A/v^s_A(T)\;.
\end{eqnarray}
\end{mathletters}
Replacing the renormalization coefficient $R_A(T)$ by its value at the
critical point, $R_A^c=1$,
one reproduces the leading-order result, Eq.~(\ref{f-sqrt}).
However,
within the temperature interval considered, the smooth drift of
$(1-f^{s,c}_A)^{\text{eff}}$ overwhelms the small variation of the
$\sqrt{T_c-T}$ term. This is demonstrated in Figs.~\ref{fig-fs}(b,c)
by the dashed lines.
The numerically found circles are well described by this line. One
concludes that the drifting coupling coefficient $v_A^s(T)$ is
responsible for the
deviations from the leading-order asymptotics. Unfortunately,
the $v^s_A(T)$ are not available directly from experiments.

\subsection{Scaling of the $\beta$-relaxation minima}

Figure~\ref{betasc-all} shows the susceptibility master spectrum $\hat\chi$ for
$\lambda=0.75$ and $\delta=0$
as dashed curves.  The upper set of solid lines in this figure
are the spectra of the first correlator rescaled, according to
Eq.~(\ref{susc-scaling}), as
$\omega\phi''(\omega)/(\sqrt{|\sigma|}h)$.
Asymptotic validity of scaling is demonstrated: the
window of rescaled frequencies $\hat\omega=\omega t_\sigma$,
for which the rescaled
spectra are close to the master spectrum $\hat\chi$, expands with
decreasing $(T-T_c)$. Convincing agreement between $\hat\chi$ and the
$180\,\mathrm K$ result can be found as long as hopping effects
are ignored. For higher temperatures, where
$|\varepsilon|=|T-T_c|/T_c\ge0.06$, strong deviations are found.
The $T=210\,\mathrm K$ spectrum, for which $|\varepsilon|=0.17$, does not
even show a minimum. The demonstrated deviations from the scaling laws are
similar to what was explained in
Ref.~\cite{Franosch97} for the MCT solutions for the hard-sphere system.

Preasymptotic-correction effects for the variables discussed for PC
in Figs.~\ref{fit-ls-de} and \ref{fit-ns} differ from those for
the auxiliary correlator $\phi$. This is demonstrated in the lower part of
Fig.~\ref{betasc-all} for $\varepsilon=-0.17$. Deviations of the rescaled
spectra $\chi''_A(\omega)/(\sqrt{\sigma}h_A)$ from the master spectrum
$\hat\chi(\omega t_\sigma)$ are larger for the dielectric loss than for the
light scattering, and the latter are larger than those for the
neutron-scattering results. While the predicted probe independence of
$\chi''_A(\omega)/h_A$ holds rather well for $\omega<\omega_{\text{min}}$,
deviations from the factorization theorem are observed mainly for higher
frequencies. As discussed above in connection with Eq.~(\ref{f-fs}),
the large size of $f^{s,c}_A$ leaves only a $10\%$ decay of the correlator
from the initial value unity to the plateau, and this decay is influenced
by vibrational motion. This leads to the strong disturbances of the
susceptibility spectra for $\omega>\omega_{\text{min}}$.
For the neutron-scattering data for intermediate
wave vectors, this problem is not so severe, since the critical
Lamb-M\"o{\ss}bauer factor $f^c_q$ decreases with increasing $q$.
Therefore the shape of the susceptibility minimum, which is exhibited by
the two neutron-scattering results shown in Fig.~\ref{betasc-all}, is closer
to the one of the master spectrum.

The $q$-dependence of the critical amplitude $h^s_q$ for the
incoherent-neutron-scattering spectra has been measured as a byproduct
of the test of the factorization theorem,
${\phi^s_q}''(\omega)\propto h_q^s\chi''(\omega)/\omega$.
A linear law, $h^s_q\propto q$, was found within the studied wave-vector
interval
\cite{Wuttke99}. Such a strictly linear law is not compatible with the
microscopic MCT,
which predicts that the $h_q$-versus-$q$ graph
exhibits a broad asymmetric peak near the position $q_{\text{max}}$ of
the first sharp diffraction peak of the structure factor.
For small $q$, the critical amplitude increases regularly as $h_q\propto q^2
+{\mathcal{O}}(q^4)$ \cite{Fuchs98}, and thus $h_q$ exhibits
an inflection point for some $q<q_{\text{max}}$.
Whether the found linear $q$-dependence of the experimental values is due
to multiple-scattering effects, is not clear \cite{Wuttke99}.
Our schematic-model fit, however, suggests that, even if $h^s_q$ can be
approximated by a linear law, the $q$-dependence is not strictly linear but
rather given by some intermediate crossover around the inflection point.
Equation~(\ref{f-ampl-sjoe}) relates $h^s_q$ to the inverse of $v^s_q$;
thus a strictly linear law for $h^s_q$ would imply $v^s_q\propto1/q$. Such
result is added as a dash-dotted line in Fig.~\ref{param-vs}b, and it
shows that this is not consistent with our data analysis.
An ad-hoc expression, reflecting the crossover from the small-$q$ asymptote
through the inflection point is $h^s_q\propto q^2/\left[1+(q/q^\star)\right]$.
The resulting expression for $v^s_q$ is added in Fig.~\ref{param-vs}b as
dashed lines for two temperatures, and it provides a reasonable interpolation
of the found fit parameters. One would need measurements for wave vectors
of the order of $0.2\,{\mathrm{\AA}}^{-1}$ and less, in order to test for
the small-$q^2$ behavior predicted by the microscopic MCT.

There is a most bothersome preasymptotic-correction effect which can be seen
in the lower part of Fig.~\ref{betasc-all}: the
positions $\omega_{\text{min}}t_\sigma$ of the susceptibility minima
are not identical; and they are all larger than the asymptotic value
$\hat\omega_{\text{min}}$ which is shown by a diamond. Since the
bands of microscopic excitations of the correlator spectra
${\phi^s}''_A(\omega)$ are
located at much lower frequencies than that of the spectrum $\phi''(\omega)$,
the spectrum of the test variables cross over too quickly to the transient
to be able to develop the universal relaxation pattern for
$\omega>\omega_{\text{min}}$. As a result, $\omega_{\text{min}}$ gets an
offset to larger frequencies. Figure~\ref{rectplot} exhibits this result
as a rectification diagram. The asymptotic result is shown as a full straight
line: $\omega_{\text{min}}^{2a}=(\hat\omega_{\text{min}}/t_0)^{2a}|\sigma|
=\hat C(T-T_c)$ with the constant $\hat C=(\hat\omega_{\text{min}}/t_0)^{2a}
\cdot C/T_c\approx0.004$.
The positions of the observed minima can
still be interpolated reasonably by straight lines, shown in dashed.
However, the slopes of the dashed lines differ from those of the asymptotic
line. In a clear
violation of the asymptotic factorization theorem, the lines for different
probing variables are different. The linear interpolations lead to
intersections with the abscissa, which differ somewhat from the correct
value of $T_c$.
For the neutron-scattering data, this interpolation has been omitted
in Fig.~\ref{rectplot}, since the error bars obtained by
$q$-averaging do not allow for a well-determined estimate here.

There is also a strong temperature-dependent offset of the amplitude scale
relative to the scaling-law prediction. This is demonstrated in
Fig.~\ref{betasc-ls} for the light-scattering spectra.
All four rescaled minima $\chi_{\text{min}}/
\sqrt{\sigma}h_{\text{ls}}$ are far below $\hat\chi_{\text{min}}$, which is
indicated by a diamond. Moreover, with decreasing $(T-T_c)$, the
discrepancy between rescaled curves and expected asymptote does not
decrease, rather it increases. Such behavior is not anticipated from the
leading-order corrections to the scaling laws \cite{Franosch97,Fuchs98},
but it can be explained as a higher-order effects because of the
important role played by the temperature dependence of the coupling
coefficient $v^s_A$.
This drift can be eliminated by introducing an
effective amplitude $h^{\text{eff}}_{\text{ls}}$, as discussed above
in connection with Eq.~(\ref{f-fs-eff}).
The result is given by the upper set of curves
in Fig.~\ref{betasc-ls}. Indeed, the discrepancies between asymptotics and
rescaled curves are reduced, and they now decrease with decreasing $(T-T_c)$.
But even for $T=190\,\mathrm K$, i.~e.\ for $|\varepsilon|=0.06$, there is
a considerable offset of the minimum intensity from the scaling result.
The $180\,\mathrm K$ curve demonstrates the approach towards the
asymptotic limit, would hopping be absent.
There still is a clear deviation between rescaled spectrum
and scaling-law result, which increases with increasing $\omega$ for
$\omega>\omega_{\text{min}}$. But the sign and size of this effect are
similar to what was found for the hard-sphere system for wave vectors
yielding a plateau $f_q^c$ as large as $f^c_{\text{ls}}$
\cite{Franosch97}.

\subsection{Summary of the Scaling-Law Analysis}

It is, of course, more satisfactory to interpret data for glassy dynamics
with the set of universal formulas provided by MCT for the asymptotic
dynamics near a glass-transition singularity than to explain experimental
findings within schematic models. The more probing variables $A$ are
taken into account, the more convincing such an analysis is, since the
universal results also imply connections between spectra measured for
different $A$. The preceding work on PC
\cite{Du94,Lunkenheimer97d,Lunkenheimer97,Schneider99,Wuttke99}
exemplifies these statements. However, the data are influenced by
preasymptotic effects, and one cannot judge the relevance of these
correction effects, if one does not know the underlying microscopic MCT
equations. Forcing data into the universal formulas can thus lead to
self-contradicting results, as the preceding subsections have demonstrated.
While the spectral shapes are rather robust and the rectification
diagram for the scales
appears correct and leads to a reasonable estimation of $T_c$, as shown
by the dashed lines in Fig.~\ref{rectplot}, the prefactors for the
asymptotic formulas extracted from the data can be quite wrong. This error
cannot be noticed if one studies a single probing variable $A$ only, but it
appears as a violation of the factorization theorem if one compares
spectra for different $A$. One concludes that the problems with the analysis
discussed in Ref.~\cite{Wuttke99} are neither due to inadequate application
of MCT results nor due to failures of MCT.
Rather they reflect the properties of MCT; more precisely,
they exemplify the limitations for the application of asymptotic laws.

A general rule \cite{Franosch97,Fuchs98} for the test of the
$\beta$-relaxation-scaling law is corroborated by the present analysis:
if the nonergodicity parameter $f_A^c$ is large, i.~e.\ if the $\alpha$-peak
strength is large compared to the strength of the microscopic-excitation
peak of the susceptibility spectrum, the preasymptotic corrections are
very important. This is especially true for the discussed light-scattering
and dielectric-loss spectra. Neutron-scattering spectroscopy has the
advantage that $f_q^c$ can be shifted by changing the wave vector $q$.
Therefore, we found the scaling-law analysis to work best for the
neutron-scattering data of Ref.~\cite{Wuttke99}. It would be very informative
to corroborate this finding by a measurement of the expected
$\sqrt{T_c-T}$ anomaly of the Debye-Waller or Lamb-M\"o{\ss}bauer factor.

\section{The dielectric modulus}
\label{de-modulus}

Since the memory kernel $m^s_q(t)$
in the mode-coupling approach is expressed as
a polynomial of the density correlators, Eq.~(\ref{fullmct-m}),
this quantity
shows the same asymptotic scenario as the correlators
themselves. From the factorization theorem for the correlators
one concludes in leading order for the kernels:
$m^s_A(t)=f^{s,c}_{M,A}+h^s_{M,A}G(t)$.
Eqs.~(\ref{sjoegren-m},\ref{f-ampl-f12},\ref{f-ampl-sjoe}) determine
the plateau $f^{s,c}_{M,A}$ and the critical amplitude $h^s_{M,A}$
for the memory kernel of probing variable $A$:
\begin{eqnarray}\label{f-ampl-mod}
  f^{s,c}_{M,A}=v^s_Af^cf_A^{s,c},\nonumber\\
  h^s_{M,A}=v^s_A\left(f^ch_A^s+f^{s,c}_Ah\right)=v^s_A h\;.
\end{eqnarray}
While in the above discussion $f^{s,c}_A$ was found to be larger than
$90\%$, such that the square-root singularity is suppressed to
a below-$10\%$ effect, the situation for the memory kernel is different.
The coupling coefficient $v^s_A$ now plays the role of a normalization
constant.
If one introduces the normalized memory kernel
in analogy to the normalized correlators $\phi_x(t)$,
$\hat m^s_A(t)=m^s_A(t)/v^s_A$, such that $\hat m^s_A(t\rightarrow0)=1$,
one gets for the normalized plateau:
\begin{equation}\label{f-mod-n}
  \hat f^{s,c}_{M,A}=f^cf_A^{s,c}\;.
\end{equation}
In cases where $f^{s,c}_A$ is close to unity, like in our analysis of the
dielectric-loss
and light-scattering data, one can approximate
$\hat m^s_A(t)\simeq f^c+hG(t)$. This
equals the asymptotic
expression for the first correlator.
Thus, we can expect $\beta$ scaling
for the memory kernel of the probe-variable $A$
to work equally well as for the first correlator and thus better than for
the corresponding probe-variable correlator.
Let us examine this in detail for the memory kernel of
$\phi^s_{\text{de}}$ underlying
the fit to the dielectric-susceptibility spectra.
For the light-scattering data, qualitatively the
the same picture arises.

In the upper part of Fig.~\ref{betasc-modulus}, three spectra of the
memory kernels for the dielectric function, rescaled to
$-\omega m_{\text{de}}''(\omega)/\sqrt{|\sigma|}$, are plotted as solid lines
for three temperatures above $T_c$. The asymptotic
$\beta$-susceptibility spectrum
for $\lambda=0.75$ is again shown as a dashed line. While the picture
shows some similarity to the situation found in the upper part of
Fig.~\ref{betasc-ls} for the light-scattering susceptibilities, the reason
for the deviations from the scaling law are different. This can be inferred
from the lower part of Fig.~\ref{betasc-modulus}, where the same scaling
is shown for the normalized memory functions. Here, the solid lines represent
$-\omega\hat m_{\text{de}}''(\omega)/\sqrt{|\sigma|}h$.
One notices, that the
$\alpha$-peak strength is remarkably smaller than in the dielectric
susceptibility, and comparable to that for the first correlator of the
model. Similarly, we find the standard scenario for the approach of the
rescaled spectra
to the master curve. The deviations from the asymptotics are
qualitatively the same as exhibited in the upper part of Fig.~\ref{betasc-all}
for $\omega\phi(\omega)$. Thus one concludes: the deviations from
scaling seen in the upper part of Fig.~\ref{betasc-modulus} are mainly due
to the $T$-dependent normalization $v^s_A$,
and not, as in the case discussed in
connection with Fig.~\ref{betasc-ls}, due to microscopic crossover effects.

The question of normalization becomes even clearer for the effective
non-ergodicity parameters.
Figure~\ref{f-modulus}(a) shows the unnormalized values $f^s_{M,\text{de}}$
as open circles. The full line exhibits the asymptotic prediction
$f^s_{M,\text{de}}=f^{s,c}_{M,\text{de}}+h^s_{M,\text{de}}
\sqrt{\sigma/(1-\lambda)}$ for $T<T_c$ and
$f^s_{M,\text{de}}=f^{s,c}_{M,\text{de}}$ for $T\ge T_c$.
Again, the drifting coupling coefficient
$v^s_{\text{de}}$ is responsible for masking the predicted square-root law.
But, unlike in
Fig.~\ref{fig-fs}, this is only true for the unnormalized quantity.
The normalized function $\hat f^s_{M,\text{de}}$, shown as filled circles in
Fig.~\ref{f-modulus}(b), exhibits good agreement with the asymptotic law.
As for the values discussed for the tagged-particle density correlators,
the drift of
$v^s_{\text{de}}$ still results in a temperature dependence of
$\hat f^s_{M,\text{de}}$ for $T>T_c$, but this drift is now reduced to
a $10\%$ effect. The asymptotic value of $\hat f^{s,c}_{M,\text{de}}$
differs only about $5\%$ from the one for the first correlator, $f^c$,
which is shown as a dash-dotted line in Fig.~\ref{f-modulus}(b).
It is remarkable, that even for the unnormalized quantity the position of
$T_c$ can be estimated better than it could be done for the
plateau values of the tagged-particle density correlators. This can
be done by
noticing that the slope of a linear interpolation of the data changes
when going over from $T<T_c$ to $T>T_c$.

From the MCT equations (\ref{fullmct}) with the hopping kernel set to zero,
one derives the expression for the dynamic susceptibility,
Eq.~(\ref{susceptibil}), in terms of
the memory kernel $m^s_A(z)$,
\begin{equation}\label{susc-m}
  \chi_A(z)=-{\Omega^s_A}^2\chi_A/\left[
    z^2-{\Omega^s_A}^2+z M^{\text{reg}}_A(z)
   +{\Omega^s_A}^2z m^s_A(z)
  \right]\;.
\end{equation}
Let us define a dynamical susceptibility $\chi_{M,\text{de}}(z)$
corresponding to
the kernel $m^s_{\text{de}}(t)$ in analogy to Eq.~(\ref{susceptibil}):
\begin{equation}\label{susceptibil-M}
  \chi_{M,\text{de}}(z)=z m^s_{\text{de}}(z)+m^s_{\text{de},0}\;,
\end{equation}
with $m^s_{\text{de},0}=m^s_{\text{de}}(t=0)$.
Then one can write for the dielectric function
$\varepsilon(z)=\varepsilon_\infty+4\pi\chi_{\text{de}}(z)$
\begin{mathletters}
\begin{equation}\label{eps1}
  \varepsilon(z)-\varepsilon_\infty=\frac{-4\pi\chi_{\text{de}}}
  {\left(z/\Omega^s_{\text{de}}\right)^2-1-m^s_{\text{de},0}
  +\left[iz\nu^s_{\text{de}}/{\Omega^s_{\text{de}}}^2
  +\chi_{M,\text{de}}(z)\right]}\;.
\end{equation}
The inverse of the dielectric function, $1/\varepsilon(z)$, is
occasionally considered as the dielectric modulus \cite{Macedo72}.
The exact Mori-Zwanzig representation, Eq.~(\ref{susc-m}), suggests to
rather consider $\left[\varepsilon(z)-\varepsilon_\infty\right]^{-1}$,
i.~e.\ $\chi_{\text{de}}^{-1}(z)$. This function consists of a quadratic
polynomial in the frequency, $\left(z/\Omega^s_{\text{de}}\right)^2-1
-m^s_{\text{de},0}$,
a white-noise background, $iz\nu^s_{\text{de}}$, and a non-trivial
part $\chi_{M,\text{de}}(z)$. The latter has all the standard properties
of a susceptibility, in particular it obeys Kramers-Kronig relations.
There is the trivial relation between the spectrum
$\chi_{M,\text{de}}''(\omega)$ and the dielectric function
\begin{equation}\label{modulus}
  \mathrm{Im}\left[\left[\hat\varepsilon-\varepsilon(\omega)\right]^{-1}
  \right]=\frac{1}{4\pi\chi_{\text{de}}}
  \cdot\left[\omega\left(\nu^s_{\text{de}}/{\Omega^s_{\text{de}}}^2\right)
  +\chi_{M,\text{de}}''(\omega)\right]\;.
\end{equation}
\end{mathletters}
Here, $\varepsilon_\infty$ was replaced by the constant $\hat\varepsilon$,
discussed above in connection with the fit of $\varepsilon'(\omega)$.

The full lines in Fig.~\ref{mem-de-cmp} exhibit the right-hand side of
Eq.~(\ref{modulus}), evaluated with the model parameters used for the
interpretation of the dielectric-loss spectra in Fig.~\ref{fit-ls-de}.
The symbols exhibit the left-hand side of Eq.~(\ref{modulus}) calculated
with the data from Ref.~\cite{Schneider99} and $\hat\varepsilon$
determined in connection with the fits of $\varepsilon'(\omega)$
in Fig.~\ref{fit-de.re}.
Figure~\ref{mem-de-cmp} shows that the fit is of equal quality as the
ones shown for the direct analysis of the dielectric loss spectra. However,
to produce the result in Fig.~\ref{mem-de-cmp}, one has to be careful to
subtract the right value
of $\hat\varepsilon$. The error bars shown in the figure for
$T=253\,\mathrm K$ indicate the influence of subtracting $\hat\varepsilon\pm1$
instead of $\hat\varepsilon$ to estimate the uncertainty introduced by this
procedure. One notices that the shape of the curves for the high-frequency
part is influenced. Thus an analysis based on
$\left[\hat\varepsilon-\varepsilon(\omega)\right]^{-1}$ is
only practicable, if one can avoid these problems of the inversion
procedure.
Up to trivial terms, the left-hand side of Eq.~(\ref{modulus}) is identical
with the spectrum $\omega m_{\text{de}}''(\omega)$ discussed in
Fig.~\ref{betasc-modulus}. This quantity can be explained well by the
$\beta$-relaxation-scaling laws. Thus for PC, the corrections to the
asymptotic laws are smaller for $\left(\left[\varepsilon(\omega)
-\hat\varepsilon\right]^{-1}\right)''$ than for $\varepsilon''(\omega)$.
In particular it is shown by Fig.~\ref{f-modulus} that the square-root
singularity can be identified from a discussion of
$\chi_{M,\text{de}}(\omega)$.

\section{Conclusion}\label{conclusions}

It was exemplified for propylene carbonate as a typical glass-forming
van-der-Waals system,
that the susceptibility spectra measured by three different
experimental techniques can be described well by a schematic MCT model.
Several decades of intensity change in the GHz frequency window as seen
in light-scattering and dielectric-loss experiments with a sensitive
temperature dependence typical for glass-forming liquids are fitted. Also,
the results from incoherent neutron scattering, probing the dynamics for
different wave vectors could be included in this simultaneous fit.
Real-part data from the dielectric experiment have been successfully
analyzed as well
to further corroborate the consistency of the schematic-model fit.
For temperatures ranging from the critical value
$T_c\approx180\,\mathrm K$ to well above the
melting point, the
range of applicability of the model includes both $\alpha$- and
$\beta$-relaxation windows, as well as the crossover to the microscopic
spectrum. Below $T_c$ and down to the glass-transition temperature
$T_g$, a rather simple approach to
account for hopping phenomena improves the fit for the
$\beta$-minimum regime, but fails to describe the $\alpha$-peak below $T_c$.

The schematic model used in the fit
captures the general features of the glass-transition scenario
predicted by the full microscopic theory. Still, it allows to go further
than an analysis based on the asymptotic predictions of MCT only. In
particular, we have used the schematic model to investigate which features
of the measured spectra can be described by asymptotic laws, and
where preasymptotic corrections set in. We find
that the asymptotic formulas qualitatively give an adequate description of the
data.
Thereby the preceding studies
\cite{Du94,Ma96,Lunkenheimer97d,Lunkenheimer97,Schneider99,Boerjesson90,Wuttke99}
are corroborated. But we demonstrated also, that preasymptotic-correction
effects cause important quantitative differences between the data and
the scaling-law results.
One aspect of this is the $T$-drift of the critical amplitude $h_{\text{ls}}^s$
noted in an earlier analysis of PC light-scattering
data \cite{Du94}. The drift of the
coupling constant $v_{\text{ls}}^s$
is not sufficient to explain this, as was demonstrated
in Figs.~\ref{beta-tdom} and \ref{betasc-ls}.
Also, the crossover to the microscopic excitations
influences the height of the spectra at the $\beta$ minimum.
For the measurements analyzed, scaling works best with the neutron-scattering
data, due to the relatively low plateau values $f^{s,c}_q$.
An asymptotic analysis of the dielectric modulus could even work better in this
respect. But due to uncertainties in the inversion of the dielectric
function, such analysis is not practicable unless the modulus itself is
measured directly.

\acknowledgments{We thank H.~Z.\ Cummins, M.\ Fuchs,
P.\ Lunkenheimer, M.~R.\ Mayr, U.\ Schneider, A.~P.\ Singh, and J.\ Wuttke
for many helpful discussions and the authors of Ref.\ \cite{Du94},
Refs.\ \cite{Lunkenheimer97d,Lunkenheimer97,Schneider99}, and
Ref.\ \cite{Wuttke99} for providing us with their files for
the various propylene-carbonate data.
This work was supported by Verbundprojekt BMBF 03-G05TUM.}

\newpage

\begin{figure}
\caption{\label{fit-ls-de}
  Susceptibility spectra for propylene carbonate (PC,
  $T_m\approx218\,\mathrm K$, $T_g=160\,\mathrm K$) as measured by
  depolarized-light-scattering (upper panel, data from
  Ref.~\protect\cite{Du94}),
  and dielectric-loss spectroscopy (lower panel, data from
  Ref.~\protect\cite{Schneider99}),
  normalized with a temperature-independent
  static susceptibility. Temperatures are in
  steps of $10\,\mathrm K$, unless indicated otherwise; in the dielectric
  measurement, $T=243$, $263$, $273$, $283\,\mathrm K$ are missing, for the
  light-scattering experiment, the highest temperature is $250\,\mathrm K$
  (see text for details). The full lines are fits by solutions of the
  two-component
  schematic MCT model defined in Sec.~\protect\ref{model-def} with parameters
  as described in the text. The dashed lines indicate a white-noise
  spectrum, $\chi_{\text{wh.n.}}''(\omega)\propto\omega$;
  the dash-dotted line in the upper panel exhibits the
  asymptote of the critical spectrum $\chi''(\omega)\propto\omega^a$
  according to Eq.~(\protect\ref{crit-omega}),
  with $a=0.30$ corresponding to the value of $\lambda=0.75$. The dotted
  line shows the solution of the model for $T=T_c$ and
  hopping terms neglected.
}
\end{figure}

\begin{figure}
\caption{\label{fit-ns}
  Susceptibility spectra for PC as measured by incoherent
  neutron scattering for three wave vectors $q$,
  from Ref.~\protect\cite{Wuttke99}.
  Temperatures are $T=210$, $220$, $230$, $240$, $251$, $260$,
  $285\,\mathrm K$,
  where alternating open and filled symbols have been used
  to help distinguishing different data sets. For lower temperatures,
  data points below $10\,\mathrm{GHz}$ are seriously affected by the
  spectrometer's resolution function and therefore not shown.
  Full lines are fits as in Fig.~\protect\ref{fit-ls-de}.
}
\end{figure}

\begin{figure}
\caption{\label{path}
  Vertices $v_1$, $v_2$ for the first mode-coupling functional,
  Eq.~(\protect\ref{f12-m}), used for the fits shown in
  Figs.~\protect\ref{fit-ls-de} and \protect\ref{fit-ns}. In the above diagram,
  the thick line represents the curve of glass-transition singularities,
  Eq.~(\ref{f12-lambda}),
  while the thin line serves as a guide to the eye indicating the chosen
  path. Each dot corresponds to one temperature. The lower diagram
  shows the separation parameter $\sigma$,
  Eq.~(\ref{sigma-f12}),
  as a function of $T$; the critical temperature $T_c\approx180\,\mathrm K$
  is determined from the zero of the shown regression line.
}
\end{figure}

\begin{figure}
\caption{\label{param-vs}
  (a) Coupling coefficients $v^s_A$ for the second mode-coupling functional,
  Eq.~(\protect\ref{sjoegren-m}), used for
  the fits shown in Figs.~\protect\ref{fit-ls-de} and \protect\ref{fit-ns}
  as functions of temperature. The squares refer to the light-scattering data
  and the circles to the dielectric-loss spectra; the lines through the
  symbols are guides to the eye. The lines without symbols connect the
  $v^s_{\text{ns}}(q)$ used for the neutron-scattering data for 10
  wave vectors $q=0.5$, $0.6$, $\ldots$, $1.4\,{\mathrm{\AA}}^{-1}$ (from
  top to bottom). The vertical dashed line indicates the critical
  temperature $T_c$.
  (b) Coefficients $v^s_{\text{ns}}(q,T)$ for the neutron-scattering data
  as functions of $q$ for various
  fixed $T$. From top to bottom, the temperatures increase from
  $T=210\,\mathrm K$ to $285\,\mathrm K$ (as given in Fig.~\ref{fit-ns}).
  The dashed lines indicate $A/q+B/q^2$ laws to visualize the difference
  to a $1/q$-law behavior, which is shown as a dot-dashed line;
  see text for details.
}
\end{figure}

\begin{figure}
\caption{\label{param-micr}
  Oscillator frequencies $\Omega^s_A$ and damping constants $\nu^s_A$ in
  $\mathrm THz$ specifying the transient motion for the second correlator as
  functions of temperature as used for the fits shown in Figs.~\ref{fit-ls-de}
  and \ref{fit-ns}. The squares refer to the light-scattering
  spectra, circles to the dielectric loss spectra; lines through the
  symbols are guides to the eye. For the neutron-scattering spectra,
  $\nu^s_{\text{ns}}$ was taken $q$-independent (diamonds in the lower
  panel), and the $\Omega^s_{\text{ns}}$ exhibit the $q\cdot\sqrt{T}$-behavior
  shown in the upper panel (lines without symbols, $q$ range as in
  Fig.~\ref{param-vs}). The vertical dashed lines indicate the critical
  temperature $T_c$.
}
\end{figure}

\begin{figure}
\caption{\label{param-delta}
  In the upper part, the hopping coefficient $\Delta$
  entering Eq.~(\ref{schematic-a}) for the first correlator
  used for the fits shown in Fig.~\ref{fit-ls-de} to the dielectric (circles)
  and light-scattering (squares) data is exhibited. The values follow an
  Arrhenius-type temperature dependence, indicated by a straight line.
  For the fit to the neutron-scattering data, the same values could be used,
  but show no influence on the fit curves in Fig.~\ref{fit-ns} (see text
  for details). For the fit to the dielectric data, an additional hopping
  coefficient $\Delta^s_{\text{de}}$ for the second correlator had to be
  used, shown by the circles in the lower part of this figure; here, the
  values $(\Omega^s_{\text{de}}/\Omega)^2\Delta^s_{\text{de}}$
  follow an Arrhenius law indicated by a straight line.
  The vertical dashed line indicates $1000/T_c$ with $T_c=180\,\mathrm K$.
}
\end{figure}

\begin{figure}
\caption{\label{delta-d0cmp}
  The solid lines reproduce the susceptibility spectra $\chi''_A(\omega)$
  from Fig.~\ref{fit-ls-de} used for the fit to the light-scattering (upper
  panel) and the dielectric-loss spectra (lower panel), respectively.
  Dashed lines are solutions using the same model parameters, but with
  hopping effects ignored: $\Delta=\Delta^s_{\text{de}}=0$.
  Notice that the dashed lines for temperatures $T$
  below $T_c$ exhibit a ``knee'' which is
  located between $10$ and $100\,\mathrm GHz$ and moves to higher
  frequencies with decreasing $T$.
}
\end{figure}

\begin{figure}
\caption{\label{fit-de.re}
  Measured values for the real part $\varepsilon'(\omega)$ of the dielectric
  function from Ref.~\protect\cite{Schneider99} for $T=173\,\mathrm K$
  through $T=233\,\mathrm K$ in steps of $10\,\mathrm K$, and for
  $T=253\,\mathrm K$ (from left to right).
  Experimental data have been shifted by $\hat\varepsilon$ to account for
  an unknown background; see text for details. Temperatures $T=163\,\mathrm K$
  and below, and $T=293\,\mathrm K$ are not shown in order to avoid
  overcrowding of the figure. For the same reason, the data points for
  frequencies above $1\,\mathrm{GHz}$ have been partially
  removed for all but the highest temperature and are only shown
  in the inset.
  The full lines are the real parts of the calculated susceptibilities for
  the same model parameters as used for the curves in Fig.~\ref{fit-ls-de}.
}
\end{figure}

\begin{figure}
\caption{\label{beta-tdom}
  The full lines show $\hat\phi_x(t)=(\phi_x(t)-f_x^c)/h_x$ for times where
  $\hat\phi_x(t)>0$. The
  dash-dotted straight lines exhibit $(t/t_0)^{-a}$, with
  $t_0=0.035\,\mathrm{ps}$ and $a=0.30$. The dashed lines
  show the $\beta$-correlators $G(t)$ for the indicated temperatures (see
  text). The
  uppermost set of curves refers to the first correlator of the schematic
  model. The other two sets refer to the light-scattering and the
  dielectric response studied in Fig.~\protect\ref{fit-ls-de}.
  The sets are shifted vertically by one (light scattering), respectively
  two (dielectric response) decades for clarity.
}
\end{figure}

\begin{figure}
\caption{\label{fig-fs}
  Effective nonergodicity parameters determined
  from the calculated $\phi_x$-versus-$\log t$ curves for the spectra shown in
  Fig.~\protect\ref{fit-ls-de}: $f(T)$ for the first correlator $\phi(t)$
  (open circles), and $f^s_A(T)$ for the correlators $\phi^s_{\text{ls}}(t)$
  and
  $\phi^s_{\text{de}}(t)$ (filled squares and circles, respectively).
  The solid lines exhibit the leading-order-asymptotic laws for $f_x$ from
  Eq.~(\protect\ref{f-sqrt}) for $T\le T_c$ and $f_x^c$ for $T\ge T_c$. The
  vertical dashed line indicates the crossover temperature
  $T_c=180\,\mathrm K$.
  The crosses in the lowest panel are estimations of $f^s_{\text{de}}$
  obtained in Ref.~\protect\cite{Schneider99}
  by fitting the measured $\varepsilon'(\omega)$ data with
  a Cole-Davidson function.
  The dashed lines through the solid symbols connect the values calculated from
  Eq.~(\protect\ref{f-fs-eff}) (see text for details).
}
\end{figure}

\begin{figure}
\caption{\label{betasc-all}
  The upper part of the figure (scale on right axis)
  shows the rescaled susceptibility spectra
  for four temperatures for the first correlator of the
  schematic model, $\omega\phi''(\omega)/h\sqrt{|\sigma|}$, as a function of
  $\hat\omega=\omega t_\sigma$. Hopping terms have been set to zero.
  The lower part (scale on the left axis) shows the
  $\omega{\phi^s_A}''(\omega)/h_A\sqrt{|\sigma|}$-versus-$\hat\omega$
  curves for $T=210\,\mathrm K$ for the light-scattering
  (line with squares) and neutron-scattering correlators at $q=1.0\,
  {\mathrm{\AA}}^{-1}$ and $q=1.1\,{\mathrm{\AA}}^{-1}$ (lines), and for
  $T=213\,\mathrm K$ for the dielectric correlators (line with circles).
  The dashed lines indicate the master spectrum
  for $\lambda=0.75$ with the minimum position $(\hat\omega_{\text{min}},
  \hat\chi_{\text{min}})$ marked by a diamond.
}
\end{figure}

\begin{figure}
\caption{\label{rectplot}
  Rectification diagram for the minimum positions $\omega_{\text{min}}$ of
  the dielectric-loss (circles),
  light-scattering (squares), and neutron-scattering (triangles) fit. The
  neutron-scattering data have been averaged over the investigated $q$ range,
  and error bars indicate the smallest and highest values.
  The vertical dashed line marks the critical point $T_c$. The solid line is
  the asymptotic-formula result for the minimum positions calculated from
  the $\beta$-relaxation-scaling law for the fit parameters of the model
  used in Figs.~\protect\ref{fit-ls-de} and \protect\ref{fit-ns}. The
  dashed lines show the linear interpolation through the data points,
  restricted to the regime $190\,{\mathrm K}\le T\le250\,{\mathrm K}$, for
  the dielectric-loss and for the light-scattering data, respectively.
}
\end{figure}

\begin{figure}
\caption{\label{betasc-ls}
  Scaling-law analysis of the MCT solutions shown
  in Fig.~\protect\ref{fit-ls-de} for the fit of
  the light-scattering spectra. The lower set of curves
  shows the rescaled spectra $\chi''_{\text{ls}}(\omega)/h_{\text{ls}}
  \sqrt{|\sigma|}$ for $T=220$, $210$, $200$, $190\,\mathrm K$ (from top to
  bottom). The upper set of curves shows the same spectra, but now rescaled
  with $h_{\text{ls}}$ replaced by the effective amplitude
  $h_{\text{ls}}^{\text{eff}}$ from Eq.~(\protect\ref{f-fs-eff}). Notice that
  the order of the rescaled spectra is now inverted.
  The asymptotic result $\hat\chi$ is plotted as a dashed line with the
  minimum position marked by a diamond. The
  result for $T=180\,\mathrm K$ with hopping effects ignored is added
  to demonstrate the approach towards the scaling limit.
}
\end{figure}

\begin{figure}
\caption{\label{betasc-modulus}
  Scaling-law analysis of the spectra for the memory kernel from the
  MCT solutions shown in Fig.~\protect\ref{fit-ls-de} to interpret the
  dielectric function.
  The upper set of solid curves shows the unnormalized rescaled
  spectra, $-\omega m_{\text{de}}''(\omega)/\sqrt{|\sigma|}$,
  for temperatures $T=193$, $203$, $213\,\mathrm K$.
  In the lower part of the figure, the
  scaling is shown for the normalized rescaled spectra
  $-\omega\hat m_{\text{de}}''(\omega)/\sqrt{|\sigma|}$
  (see text for details).
  The dashed lines are the master spectra, with the minimum
  positions marked by diamonds.
}
\end{figure}

\begin{figure}
\caption{\label{f-modulus}
  (a) Unnormalized nonergodicity parameters $f^s_{M,\text{de}}$
  for the memory kernel from the fit to the
  dielectric spectra (open circles). The asymptotic prediction is plotted
  as a solid line, the vertical dashed line indicates $T_c$. (b) Same as in
  (a), but normalized values $\hat f^s_{M,\text{de}}$ (filled circles).
  The dash-dotted line indicates the critical plateau value for the first
  correlator of the schematic model, $f^c$.
}
\end{figure}

\begin{figure}
\caption{\label{mem-de-cmp}
  The lines show the dielectric-modulus spectra defined by the
  right-hand side of Eq.~(\ref{modulus}).
  The circles exhibit ${\mathrm{Im}}\left[\left[\hat\varepsilon-
  \varepsilon(\omega)\right]^{-1}\right]$ calculated from the
  dielectric-function data of Ref.~\protect\cite{Schneider99}. The curves
  and the data sets
  have been shifted vertically to avoid overlapping; temperatures and shift
  factors are ($183\,\mathrm K$, $1$),
              ($193\,\mathrm K$, $0.316$),
              ($203\,\mathrm K$, $0.1$),
              ($213\,\mathrm K$, $0.0316$),
              ($223\,\mathrm K$, $0.01$),
              ($233\,\mathrm K$, $0.00316$),
              ($253\,\mathrm K$, $0.001$).
  For the highest temperature, an estimation of the uncertainty
  introduced by inverting
  the experimental data is given as error bars; see text for details.
}
\end{figure}

\end{document}